# Comment on Papers by Foot, Kobakhidze, McDonald and Volkas Relating to Scale Invariance Symmetry

Hitoshi Nishino[1)] and Subhash Rajpoot[2)]

*Department of Physics & Astronomy*
*California State University*
*1250 Bellflower Boulevard*
*Long Beach, CA 90840*

## Abstract

We point out that the works described by Foot et al. in arXive: 0706.1829 [hep-ph] and arXive: 0709.2750 [hep-ph] are derivatives of our work described in hep-th/0403039, the extended version of which was published in *"Standard Model and SU(5) GUT with Local Scale Invariance and the Weylon"*, AIP Conf. Proc. **881** (2007) pp. 82, Melville, New York, 2006. We are wondering how many motions (and publications!) they will go through before finally admitting that they have re-discovered our model, and of course, as is the usual practice these days, claiming afterwards to the world of their independent arrival at our model. Reference to our original work is long overdue.

---

[1)] E-Mail: hnishino@csulb.edu
[2)] E-Mail: rajpoot@csulb.edu

Scale invariance symmetry was revived in our work [1] and has recently become an important topic of research. In the past, we have commented in [2] and [3] on works appearing and published in the literature that are derivatives of our original work [1]. For more details, we refer our readers to these comments [2][3].

An important issue in elementary particle physics is to seek explanation of the origin of masses of quarks, leptons and other fields. To this end, we extended the standard model in such a way so as to retain the original standard model construct [4] as well as address the issue of the problem of mass though scale invariance.

For completeness, we again recapitulate the main ingredients of our work [1]:

(i) It has local scale invariance that was considered by Weyl many years ago for different reasons [5].
(ii) Since mass and gravitational interactions are ultimately linked via Newton's law, we included gravitational interactions in such a way that Einstein-Hilbert action follows after spontaneously breaking of scale invariance [1].
(iii) The Higgs sector is extended to include a real scalar singlet [1].
(iv) The fermion spectrum is extended to include right-handed neutrinos [1].
(v) The mass spectrum of the quarks and leptons and other fields is generated by explicit spontaneous breaking scale invariance [1].
(vi) In our work the gauge field associated with local gauge invariance is called the Weylon, named in honor of the proponent of the gauge principle. It becomes massive after scale symmetry breaking [1].
(vii) The Weylon has either no or feeble couplings to the quarks and leptons [1].
(viii) Neutrinos acquire masses via the see-saw mechanism. The see-saw scale is, of course, provided by the scale associated with the primary descent [1].

We now comment on the work by Foot et al. [6][7] in the light of our model in [1]. Foot, Kobakhidze, McDonald and Volkas take our model, and do the following:

(1) Excise local scale invariance in our model [1]. It is obvious that once scale invariance is excised, it remains as a global scale invariance in our model.
(2) Use exactly the same scalar potential as in our model [1]. (Their eqs. (3) and (4) in [7])
(3) Now they break symmetry by radiative corrections, whereas in our model [1], it is broken explicitly.
(4) As this was not enough, they sneak in the $\varphi^2 R$-term and justify the emergence of Planck scale in their work. Note that the $\varphi^2 R$-term and the Planck scale are a naturally occurring theme of our model.

The above remarks should be read in the light of Foot *et al.*'s other work [8], while again derivations of our work [1].

It is to be noted that once local scale invariance is replaced by global scale invariance in our model, one of the scalar particles is the Nambu-Goldstone boson. Foot, Kobakhidze, McDonald and Volkas [6][7][8] dwell heavily on this feature of our model [1]. They compute one-loop radiative corrections à la Coleman-Weinberg, and after a long song and dance claim to solve the hierarchy problem. However, their one-loop radiative corrections are in a non-renormalizable model.

We are surprised that with so much overlap with our work, the authors in [6][7][8] have *deliberately* failed to recognize their work as derivative of our work [1].

As is often said, "If it talks like a duck and walks like a duck, it's a duck". Unfortunately, Foot *et al.* are under the illusion that they have created a new species of birds. Their bird is our duck, however cleverly disguised it may be.

At the present time, string theory is the leading candidate for quantum theory of gravity. It is our contention that our model will be the integral part of the low-energy limit of such a theory.